# CRYPTO ANALYSIS OF THE KEY DISTRIBUTION SCHEME USING NOISE-FREE RESISTANCES

LASZLO B. KISH[1,2]

[1]*Department of Electrical and Computer Engineering, Texas A&M University,*
*College Station, TX 77841-3128, USA*
*laszlokish@tamu.edu*

[2]*Óbuda University*[‡]*, Budapest, Bécsi út 96/B, Budapest, H-1034, Hungary*

**Abstract:** Known key exchange schemes offering information-theoretic (unconditional) security are complex and costly to implement. Nonetheless, they remain the only known methods for achieving unconditional security in key exchange. Therefore, the explorations for simpler solutions for information-theoretic security are highly justified. Lin et al. [1] proposed an interesting hardware key distribution scheme that utilizes thermal-noise-free resistances and DC voltages.

A crypto analysis of this system is presented. It is shown that, if Eve gains access to the initial shared secret at any time in the past or future, she can successfully crack all the generated keys in the past and future, even retroactively, using passively obtained and recorded voltages and currents. Therefore, the scheme is not a secure key exchanger, but it is rather a key expander with no more information entropy than the originally shared secret at the beginning.

We also point out that the proposed defense methods against active attacks do not function when the original shared secret is compromised because then the communication cannot be efficiently authenticated. However, they do work when an unconditionally secure key exchanger is applied to enable the authenticated communication protocol.

## 1. Introduction

An interesting hardware scheme, based on *noiseless* resistors, for secure key exchange was proposed by Lin, Ivanov, Johnson and Khatri (LIJK) [1]. The LIJK sheme [1] employs a shared random secret, as a resistor, which enables information transfer using (other) resistors, DC voltage sources, and switches. The LIJK protocol is, by a clever trick, an enhancement of a *non-functioning predecessor* of the Kirchhoff-Law-Johnson-Noise (KLJN) scheme [2-10] (the problems with the predecessor scheme are discussed in [4,7]).

---

[‡] Honorary faculty.

In this paper, we present a cryptographic analysis of the LIJK system [1] and show some of its limitations and security vulnerabilities.

First for the benefit of the Reader, we summarize some of the elements of cryptography that are necessary for this analysis.

*1.1 Secure key exchange*

In symmetric-key secure communications, Alice and Bob are using identical ciphers to encrypt and decrypt the messages [10,11]. The ciphers utilize the shared secure key (identical strings of random bits) to carry out these tasks. The process of securely sharing the key between Alice and Bob is called *secure key exchange*, or *secure key distribution*. This task is particularly difficult because it is itself a secure communication, when a cipher cannot be used, as there is no shared key yet. Thus, finding the best secure key exchange protocols is at the front of security research.

*1.2. Information-theoretic (unconditional) security*

Shared keys are typically obtained by number theoretical protocols that involve two communicators (Alice and Bob) exchanging data. The security of these protocols relies on the *assumption* that it is computationally hard (but not impossible) to extract the key from the data exchange between Alice and Bob. However, this assumption has no mathematical proof; it is only based on the common intuitive opinions of many mathematicians. Effective algorithms utilizing polynomial computing power may still exist and someone may find them. Consequently, these protocols are termed conditionally secure, contingent upon the validity of the assumption that breaking them indeed requires exponential computational power. Moreover, some of these protocols could be broken by a quantum computer, if such a device becomes available. Conditional security is *not future proof*, because an eavesdropper (Eve) can crack the key with enough time and/or computational power. From an information-theoretic perspective, the key has zero information entropy for Eve, which means *zero security*.

In contrast, *information-theoretic security* or *unconditional security* [11,12] means that the key has maximum entropy for Eve, regardless of her computational power or physical limitations. During an arbitrary attack, for a key length of $N$ bits, Eve's information entropy remains $N$ bits, which means the key is secure even if Eve has unlimited resources that are restricted by only the laws of physics. Time and cost are irrelevant.

Ciphers that are utilizing the secret keys can be made unconditionally secure without hardware components, for example, Shannon's One Time Pad (Vernon cypher) [11].

However, secure key exchange is different. Currently there are only two unconditionally secure key exchanger families:

- Quantum Key Distribution (QKD) [12-21], which is based on the *No-cloning theorem* of quantum informatics and/or *quantum entanglement*.

- The KLJN scheme [2-10] that is based on the *Second law of thermodynamics*.

Both QKD and the KLJN scheme are complex and costly to implement. Nonetheless, they remain the only known methods for achieving unconditional security in key exchange. This renders them indispensable for applications where the key's unconditional security is paramount. Therefore, the pursuit of a simpler solution for information-theoretic security, as exemplified by LIJK's efforts [1], is highly warranted.

*1.3 Authenticated communication*

Authenticated communication is used to verify the identity of the communicating parties. The communication is *not encrypted* and is available for the public. It is needed for both the QKD and KLJN schemes, in the public channel between Alice and Bob, for basic function (QKD) and/or to secure the systems against active attacks (QKD, KLJN). A small, $O[\log(N)]$, part of the secure key is used up for the digital signatures, which must be encrypted to defend against active attacks.

*1.4 Kerckhoffs's principle (Shannon's maxim)*

To state that a communication system is unconditionally secure, it must satisfy the Kerckhoffs's principle (Shannon's maxim) [11], which means the adversaries know the system and the detailed protocol, except the secure key. Any component that is a stationary component of the system and itself the protocol are assumed to be known by the adversaries. Therefore, keys must be spontaneously generated, used and annihilated.

*1.5 Secure key expansion*

Secure key expansion [22] starts with an *N* bit long secure key and produces a larger key with *M>N* bits, in a *deterministic* fashion, by a deterministic algorithm that both Alice and Bob are using. Due to the determinism and the Kerckhoffs's principle, the information entropy of the *M* bit long key cannot be larger than *N* bits. In a secure key expander, if Eve learns the original key, she can extract all the future keys.

*1.6 Shared secret in QKD and KLJN*

While a shared secure key is inherently a shared secret, certain secure key exchange protocols, such as QKD and the KLJN system, necessitate a shared secret even for the initial key exchange, when no key has been exchanged yet. This shared secret serves a dual purpose: it facilitates the authentication component of the protocol (QKD) and protects against active attacks during the first key exchange (QKD, KLJN). If no initial shared secret exists, alternative security measures must be implemented for the first run.

From the second key exchange onward, Alice and Bob can utilize a portion of the newly generated key for data authentication, rendering the original shared secret obsolete. The first key exchange, in essence, functions as a secure key expansion protocol. If Eve possesses the original shared secret, she can launch an active attack and extract the first key, enabling her to compromise future keys as well. However, if the original shared secret remains intact until the initial shared key is established, Alice and Bob can derive subsequent shared secrets for authentication from the first exchanged key.

After the first key exchange, even if Eve acquires the original shared secret later, she cannot retroactively launch a successful active attack. Consequently, the protocol continues to generate new entropy. From the second run onwards, the QKD and KLJN systems transition from key expanders to key exchangers.

## 2. The LIJK scheme

*2.1 The LIJK protocol*

The LIJK protocol [1] contains six resistors and four voltage generators. However, the published system has unnecessary redundancies and a more complicated protocol than necessary. Two of the resistors and the related steps of operation unnecessarily complicate the protocol, without any advantage. Thus we present a simplified, more hardware-effective version of it, with the same security properties, see Figures 1-3. Another (minor) change is that we use only two voltage generators with variable voltage levels, instead of four generators with switches, for a simplified presentation of the working principle of the scheme.

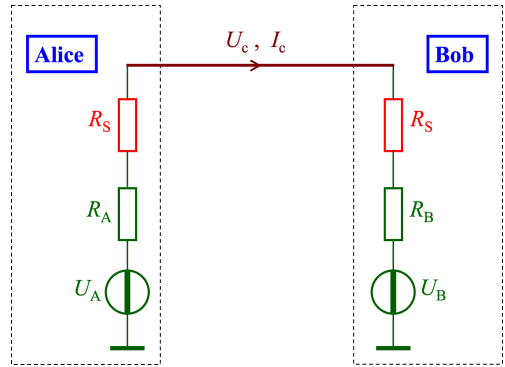

Figure 1. The simplified/essential LIJK scheme [1]. The non-functioning predecessor [4,7] of the KLJN scheme is expanded with a shared secret between Alice and Bob, which makes it working. $R_S$ is the shared secret; $R_A$, $U_A$ and $R_B$, $U_B$ are the secure resistances and voltage generators of Alice and Bob, respectively. The arrow shows the reference direction of positive current flow. All the resistors are supposed to be noise-free, meaning that the applied voltages are much greater than the thermal noise voltages of the resistors.

The essential LIJK scheme in Figure 1 shows four resistors and two tunable DC voltage sources with secret values placed in the communicating parties' (Alice and Bob) private places. The $R_S$ resistors located at both sides form a shared secret between Alice and Bob. The goal is to securely share the values of Alice's resistance $R_A$ and voltage $U_A$, and Bob's resistance $R_B$ and voltage $U_B$ between the parties. The wire line connecting Alice and Bob is public thus it is accessible for Eve's attacks.

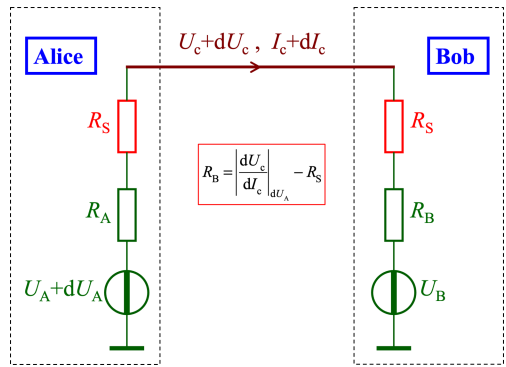

Figure 2. Extracting the secret resistance value of Bob by Alice who changes her voltage source output while measuring the public line. Eve can also monitor the line but she does not know the shared secret, so while her information entropy has decreased, its remaining value is the share secret itself.

The key exchange between Alice and Bob is straightforward. To get Bob's resistance value, Alice changes her voltage by $dU_A$ and monitors the response $dU_c$ and $dI_c$ in the wire, see Figure 2. Bob's resistance is obtained by the Kirchhoff loop law and the superposition principle:

$$R_B = \left| \frac{dU_c}{dI_c} \right|_{dU_A} - R_S .  \qquad (1)$$

Bob can act similarly, see Figure 2, and extract the secret resistance of Alice:

$$R_A = \left| \frac{dU_c}{dI_c} \right|_{dU_B} - R_S .  \qquad (2)$$

Then Alice and Bob can simply calculate the voltages $U_B$ and $U_A$ of each other, respectively, because by now they know all the resistances and their own voltages. Thus they have learned both the secrets from each other.

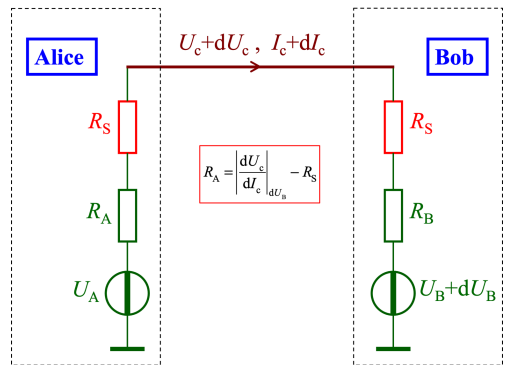

Figure 3. Bob is learning Alice's resistance with this protocol that corresponds to Alice's one, see Figure 2.

*2.2 Attacks by Eve*

Due to the Kerckhoffs's principle, Eve familiar with all the details of the protocol except the initial shared secret and the temporary random secret values. She is measuring the voltages and currents in the line and storing them for current and future use. The following considerations hold:

(i) A *passively* eavesdropping Eve can measure only the voltage $U_c$ and current $I_c$ in the cable. Thus, at the beginning (Figure 1), she can set up only two equations for the six unknown variables. Therefore, at this moment, these variables remain secret for Eve.

(ii) Nonetheless, during the key exchange between Alice and Bob, as illustrated in Figures 2, 3, and Equations 1, 2, Eve may still gain access to information that was previously unknown to her:

$$X_B = \left| \frac{dU_c}{dI_c} \right|_{dU_A} = R_S + R_B \tag{3}$$

$$X_A = \left| \frac{dU_c}{dI_c} \right|_{dU_B} = R_S + R_A \tag{4}$$

Now, Eve knows the $X$ values. If she were aware of the value $R_S$, she could extract $R_B$ by deterministic operations from Equations 4 and 5. *Thus her information entropy is only the $R_S$ value expressed in bits*.

Note: *Transient attacks* during the protocol [23] would uncover only the $X$ values thus the system is protected against them.

Furthermore, Eve can also extract the secret voltage values of Alice and Bob. She has the record of the original current and voltage in the wire, thus she can calculate:

$$U_B = U_c - X_B I_c \tag{5}$$

$$U_A = U_c + X_A I_c \tag{6}$$

Consequently, the voltages, previously considered secure in [1], are no longer secure. Therefore, the protocol can be executed with known voltages while maintaining the same level of security for the resistances.

(iii) An active (invasive) attack involving Eve injecting current or other elements into the line can be detected using Alice's and Bob's authenticated data exchange protocol and a computer model (digital twin) of the cable. Although this issue is not adequately addressed in [1], previous works have already presented a detailed description of an efficient protocol for the KLJN key exchange scheme, as seen in [24]. With authenticated data exchange and accurate cable models, the system can effectively defend against such attacks. Paper [1] discusses an active attack that malignantly modifies Alice's and Bob's keys, proposing a "challenge response" scheme as a defense mechanism. However, this approach also necessitates authenticated communication.

## 3. Conclusions

Our crypto analysis revealed the following findings:

(a) To effectively defend against active attacks, authenticated data communication using a portion of the shared secret is essential. This feature must be added to the protocol

(b) The protocol deviates from its description. Unlike QKD or KLJN, which generate new, unconditionally secure shared secrets after each execution, this protocol functions as a key expander, gradually consuming the initial shared secret. The key's information entropy remains constant throughout the protocol, even as its bit length increases. Consequently, it is classified as a key expander. If Eve ever learns the initial shared secret, she can decrypt all the keys generated by the procedure.

(c) Following this crypto analysis, it became evident that the same key expansion process can be implemented efficiently (but with the same security limitations) without any hardware requirements using an email protocol. For instance, suppose Alice and Bob share a secret, such as the number $R_S=75191$. They can generate a long series of $X_k$ values by adding random numbers $R_{A,k}$ and $R_{B,k}$ to the shared secret:

$$\{X_{A,k}\} \equiv \{R_S + R_{A,k}\} = \{75191 + R_{A,k}\}, \quad \{X_{B,k}\} \equiv \{R_S + R_{B,k}\} = \{75191 + R_{B,k}\}, \qquad (7)$$

where the $k$ index is for the $k^{th}$ exchanged key. By exchanging sufficiently large sets of generated $X_k$ values in a single email, Alice and Bob effectively execute the same protocol, eliminating hardware requirements and potentially achieving speeds superior to those of the LIJK scheme. However, more secure key expansion solutions also exist, and none of these methods increase the original entropy [22]. Therefore, these techniques cannot be considered secure key exchangers. The sole key sharing in this protocol occurs by providing the initial shared secret $R_S$.

**Acknowledgements**

The author acknowledges Dr. Sunil Khatri for stimulating discussions and expresses appreciation for his team's ingenious brainteaser, which holds both academic and educational significance.**References**

[1] P.C.K. Lin, A. Ivanov, B. Johnson, S.P. Khatri, A novel cryptographic key exchange scheme using resistors, *IEEE International Conference on Computer Design: VLSI in Computers and Processors, (ICCD)*, October 9-12, 2011, Amherst, MA, USA; *IEEE*, pp. 451-52, 2011. Electronic ISBN: 978-1-4577-1954-7, DOI: 10.1109/ICCD.2011.6081445.
[2] L.B. Kish, Totally secure classical communication utilizing Johnson (-like) noise and Kirchhoff's law, *Phys. Lett. A* **352** (2006) 178-182.
[3] A. Cho, simple noise may stymie spies without quantum weirdness, *Science* **309** (2005) 2148-2148.
[4] L.B. Kish, The Kish Cypher: The Story of KLJN for Unconditional Security, New Jersey: World Scientific, (2017).
[5] L.B. Kish and C.G. Granqvist, On the security of the Kirchhoff-law–Johnson-noise (KLJN) communicator, *Quant. Inform. Proc.* **13** (2014) (10) 2213-2219.
[6] R. Mingesz, Z. Gingl and L.B. Kish, Johnson(-like)-noise-Kirchhoff-loop based secure classical communicator characteristics, for ranges of two to two thousand kilometers, via model-line, *Phys. Lett. A* **372** (2008) 978–984.
[7] L.B. Kish, Enhanced secure key exchange systems based on the Johnson-noise scheme, *Metrol. Meas. Syst.* **20** (2013) 191-204.
[8] L.B. Kish, Protection against the man-in-the-middle-attack for the Kirchhoff-loop-Johnson (-like)-Noise cipher and expansion by voltage-based security, *Fluct. Noise Lett.* **6** (2006) L57-L63.
[9] S. Ferdous and L.B. Kish, Transient attacks against the Kirchhoff–Law–Johnson–Noise (KLJN) secure key exchanger, *Appl. Phys. Lett.* **122**, 143503 (2023).
[10] C. Chamon and L.B. Kish, Perspective—On the thermodynamics of perfect unconditional security, *Appl. Phys. Lett.* **119**, 010501 (2021).
[11] C.E. Shannon, Communication theory of secrecy systems, *Bell Systems Technical Journal* **28** (1949) 656–715.
[12] Y. Liang, H.V. Poor and S. Shamai, Information theoretic security, *Foundations Trends Commun. Inform. Theory* **5** (2008) 355–580.
[13] H.P. Yuen, Security of quantum key distribution, *IEEE Access* **4** (2016) 7403842.
[14] C.H. Bennett and G. Brassard, Quantum Cryptography: Public Key Distribution and Coin Tossing, Proc. IEEE Int. Conf. Comp., Syst., Signal 277, Process. 1, 175–179 (1984).
[15] S. Sajeed, A. Huang, S. Sun, F. Xu, V. Makarov, and M. Curty, Insecurity of detector-device-independent quantum key distribution, *Phys. Rev. Lett.* **117** (2016)9